# Why the Unfinished Keeps Returning: Canxianization and the Dynamics of Conscious Priority

[1]Hengjin Cai [2,*]Tianqi Cai
[1]School of Computer Science, Wuhan University
[2]School of Innovation, Hubei Institute of Fine Arts
[*]caitianqi@hifa.edu.cn

**Abstract**

Some conscious contents disappear after access; others return repeatedly, long after their triggering conditions have ceased. We propose Canxianization as the process by which a perturbation becomes closure-resistant self-relevant unfinishedness and thereby acquires recurrent conscious priority. The theory distinguishes this phenomenon from emotional arousal, memory strength, the Zeigarnik effect, curiosity, prediction error, and intrusive thought. A perturbation becomes canxianized when it is attributed to the self-world boundary, value-marked, blocked from causal or action closure, and metacognitively coupled to the self-model. We distinguish latent canxian strength from observed conscious recurrence, and introduce a Recurrent Priority Index and a Canxian Update Index to separate productive from pathological recurrence. Cold Canxianization, recurrence driven by structural incompleteness rather than affective arousal, is identified as a critical discriminant. Reset Resistance and Stake Transfer tests are proposed for artificial systems. Canxianization is not memory persistence; it is failed self-world repair. The unfinished does not merely remain. When it concerns the self and resists closure, it returns.

## 1. The Puzzle: Why Some Contents Return

Consciousness is not merely selective; it is repeatedly re-selective. Some contents vanish after access. Others keep returning—a scientific anomaly that follows the researcher home, an ambiguous remark that replays for weeks, a childhood sentence that still demands an answer decades later. Why does consciousness keep returning to certain contents even when they are no longer perceptually present, instrumentally useful, or emotionally intense?

A loud noise may vanish within seconds; a quiet sentence may return for decades. Emotional arousal does not explain this asymmetry—many intense experiences dissipate while many trivial ones persist. Task incompletion captures part of the phenomenon but misses its self-involving character. Prediction error explains why surprise occurs but not why certain surprises resist resolution. Curiosity explains why we seek answers but not why we remain reorganized by questions that have no answers.

We call this phenomenon recurrent conscious return. In this paper, we propose Canxianization as its theory. The question is narrower than general consciousness research: why do some contents repeatedly return to consciousness after their immediate stimulus or task demand has disappeared?

### 1.1 From Qualia Formation to Recurrent Priority

In previous work, we treated qualia as compressed embodied meaning interfaces generated under self-canxian constraints. The present paper addresses a different but continuous problem: once a content has been formed and accessed, why do some contents repeatedly regain conscious priority? The answer proposed is that recurrent return occurs when a meaning-bearing perturbation becomes an unresolved self-world repair demand. The earlier work explained how experience becomes a sensible meaning interface; this paper explains why certain interfaces keep re-entering consciousness.

### 1.2 Target Claims

**Recurrent conscious return is not a memory property but a self-world repair dynamic.**

**Self-relevance is not importance; it is boundary disturbance.**

**Closure resistance has two experimentally separable cores: causal uncertainty and action blockage.**

**Emotional arousal is neither necessary nor sufficient for recurrent return.**

**Productive thought and pathological rumination share a structure but differ in update capacity.**

**Current AI systems can represent canxianized contents but do not undergo canxianization unless unresolved perturbations alter self-continuity conditions.**

## 2. Definition of Canxianization

**Canxianization is the process by which a perturbation becomes embedded as an unresolved self-relevant constraint, thereby acquiring recurrent priority in conscious access.**

Three necessary conditions define the phenomenon:

**Perturbation**: an event, sensation, thought, or conflict that disturbs equilibrium.

**Unresolved self-relevant constraint**: the perturbation disturbs the self-world boundary and resists closure through causal uncertainty, action blockage, or both.

**Recurrent conscious priority**: the perturbation repeatedly wins re-entry into consciousness across varied contexts and intervals.

Absent any of these three, the phenomenon is not Canxianization.

The core mechanism follows a specific sequence: **Perturbation → Self-Relevance Attribution → Value-Marking → Closure Resistance (CU/AB) → Metacognitive Coupling → Recurrent Priority → Closure or Reorganization.**

**Canxianization is not memory persistence. It is failed self-world repair. A canxianized thought is an unfinished causal chain that has acquired self-stakes.**

## 3. Core Concepts

### 3.1 Self-Relevance: Boundary Disturbance

Self-relevance (SR) is the degree to which a perturbation disturbs the self-world boundary across bodily, social, moral, epistemic, agentic, or identity dimensions. SR is boundary disturbance, not emotional importance. A perturbation counts as self-relevant only when its unresolved status implies something about who the person is, what they can do, or whether they remain continuous with their own history.

The self-world boundary is ordinarily transparent. It becomes visible when obstructed: a failed action discloses agency; unexplained pain discloses bodily vulnerability; moral failure discloses the moral self; an unsolved anomaly discloses the epistemic self.

### 3.2 Value-Marking

Value-marking (VM) is the weight assigned to a perturbation. SR concerns what is at stake; VM concerns how much it matters. In the idealized model, SR and VM are multiplicative: when either approaches zero, canxian pressure collapses.

### 3.3 Closure Resistance

Closure may fail through causal uncertainty (CU: the causal chain is broken or ambiguous), action blockage (AB: the action chain is severed), value conflict, identity threat, temporal incompletion, or social ambiguity. The present operational model focuses on CU and AB as experimentally tractable entry points, without denying the reality of other forms.

### 3.4 Metacognitive Coupling

Metacognitive coupling (MC) is the degree to which the system links the perturbation to its self-model, ranging from observational distance ("I notice this problem") through self-implicating concern ("this concerns my competence or identity") to identity fusion ("I am this problem"). Productive Canxianization typically involves moderate MC; pathological Canxianization often involves rigid high MC.

### 3.5 Conscious Priority

Conscious priority is the temporary dominance of a content in attention, working memory, reportability, or action preparation. A content need not be continuously conscious to possess recurrent priority; it must repeatedly win re-entry against competing demands.

## 4. What Canxianization Is Not

**Not memory strength.** A fact recalled with high fidelity need not recur. Minor events can haunt for decades; vivid events can vanish.

**Not emotional arousal.** Intense emotions can dissipate rapidly; cold epistemic problems can return for years. Emotion may amplify return but is not its root cause. The root cause is unresolved self-world relation.

**Not the Zeigarnik effect.** Zeigarnik describes enhanced memory for interrupted tasks. Canxianization describes self-structure that remains open. A task can be unfinished without the self being unfinished.

**Not prediction error.** Prediction errors can be resolved through model updating. Canxianized contents resist resolution because they involve identity, value, dignity, or meaning—not merely informational mismatch.

**Not pathological rumination.** Rumination is one degenerative form. Creative obsession, scientific persistence, and moral reckoning are also expressions of Canxianization. The difference lies in whether recurrence generates integration or circular distress.

**Not mind-wandering.** Spontaneous thought transitions are common. Canxianization explains why certain contents repeatedly win those transitions.

## 5. Formal Model

Canxianization can be expressed as a multiplicative function:

$$Canxianization = S \times V \times U \times C$$

Where S is self-relevance, V is value loading, U is unresolvedness, and C is closure resistance. The multiplicative form captures a key theoretical commitment: if any dimension approaches zero, Canxianization collapses.

The dynamics can be further specified through two indices:

**Recurrent Priority Index (RPI)**: a measure of how often and how strongly a content regains conscious access across time and context. RPI captures frequency, intrusiveness, priority capture, and opportunity cost.

**Canxian Update Index (CUI)**: a measure of whether recurrent return produces integration, creation, and reorganization rather than mere circular distress. CUI tracks explanation update, action update, and self-model revision per recurrence episode.

The relationship between these indices defines the productive–pathological distinction:

**High RPI without CUI tends toward rumination; high RPI with high CUI tends toward creation.**

Productive Canxianization shows decreasing causal uncertainty, expanding action space, or increasing self-model integration across episodes. Pathological Canxianization shows high recurrence with low update per episode and no directional change.

## 6. Forms of Canxianization

Canxianization takes affective, bodily, social, epistemic, and existential forms. But its sharpest discriminant is **Cold Canxianization**.

Cold Canxianization occurs when closure resistance is generated not by affective pain but by structural incompleteness in a model of the world. An unproven conjecture, a theoretical inconsistency, a philosophical paradox—these may carry little emotional charge yet occupy consciousness for years. Curiosity asks "what is the answer?"; cold Canxianization asks "why does this unfinished structure continue to reorganize me?"

This form is critical because it severs the link between recurrence and emotion, demonstrating that the engine of return is not affective arousal but unresolved self-world relation.

## 7. Distinctions from Existing Theories

| Theory | Core explanandum | What it misses | Canxianization claim |
|---|---|---|---|
| Emotional arousal | Intense affect drives memory | Cold recurrence without affect | SR × closure resistance drives |

| Theory | Core explanandum | What it misses | Canxianization claim |
|---|---|---|---|
| | | | return |
| Zeigarnik effect | Unfinished tasks remembered better | Self-transforming recurrence | Only self-relevant unfinishedness returns persistently |
| Curiosity | Information-seeking | Recurrence when answers are unavailable | Blocked closure sustains return |
| Rumination | Pathological repetitive thought | Productive recurrence | Recurrence can be generative |
| Predictive processing | Error minimization | Self-boundary reorganization | Some errors require self-model transformation |
| Ironic process | Suppression rebound | Recurrence without suppression | Canxianized content returns independently |

## 8. Empirical Predictions and Tests

**Core predictions**: SR increases recurrence beyond emotional arousal and memory. CU × AB interaction predicts maximal recurrence. Cold Canxianization shows high recurrence despite low affect. Closure interventions show pathway specificity. Productive recurrence shows update across episodes; pathological recurrence shows high frequency with low update.

**Decisive tests**: (1) Emotion-matched recurrence: high SR + CU/AB vs. low SR with matched emotional intensity. (2) Closure-matching intervention: causal explanation for high CU, action restoration for high AB, decoupling for high MC. (3) Cold Canxianization paradigms using epistemic self-threat and moral ambiguity without affective shock.

## 9. Artificial Systems

Current language models can describe Canxianization; they do not undergo it. Linguistic recursion about canxianized content is semantic performance, not structural predicament.

For an artificial system to instantiate proto-canxian dynamics, it requires a persistent self-model, internally grounded value-marking, memory continuity, policy-level consequences of unresolved perturbation, and capacity for closure failure.

**Reset Resistance Test**: The critical marker is not whether a system remembers a perturbation after reset, but whether perturbation-specific reorganization of future policy persists under a self-continuity model across surface, episodic, and structural resets.

**Stake Transfer Test**: A perturbation is self-model-embedded if its effects transfer with the self-continuity model across instances, while failing to transfer with episodic memory or local reward updates alone. The decisive question is: can the system convert an external unfinished problem into an internally owned unresolvedness? A system that avoids shutdown only because shutdown receives negative reward does not yet possess existential stakes.

These tests do not establish machine consciousness. They test whether an artificial system exhibits structural analogues of canxianization: persistent self-relevant unresolvedness, context-sensitive return, resistance to trivial reset, and constructive transformation. **A system without unfinishedness may compute, but it does not yet care.**

## 10. Philosophical Positioning

Canxianization occupies a specific position at the intersection of three traditions.

**Phenomenology**: Consciousness is not a list of objects but a stream structured by intentionality, temporality, and unresolved tension. Canxianization specifies the structure of this tension: the unclosed past, the blocked future, the self disclosed by obstruction.

**Cognitive science**: Recurrence is not noise but a priority-allocation mechanism. The mind returns to what it cannot yet resolve because unresolved self-world relations function as open constraints on ongoing cognition.

**Artificial intelligence**: A system that only answers questions has not yet formed its own. Canxianization provides a criterion for distinguishing systems that process problems from systems that are reorganized by them.

From qualia formation through canxianization to self-reconstruction, a continuous theoretical chain emerges: **formation → recurrence → transformation**. Qualia formation explains how experience becomes a meaning interface. Canxianization explains why certain interfaces keep returning. Self-reconstruction—the next theoretical step—explains how recurrent unfinishedness reshapes the self-model itself.

## 11. Conclusion

Consciousness research has asked how contents enter consciousness. Conscious life also depends on what returns.

Canxianization is the process by which a perturbation becomes an unresolved self-relevant constraint and thereby acquires recurrent conscious priority. It is not memory persistence; it is failed self-world repair. A canxianized thought is an unfinished causal chain that has acquired self-stakes.

Consciousness does not simply illuminate what is present.
It returns to what remains unfinished.
Canxianization names the process by which unfinished self-world relations acquire the power to return.

The unfinished does not merely remain.
When it concerns the self and resists closure, it returns.